\documentstyle[12pt,moriond,psfig]{article}
\begin{document}
\heading{Spiral Galaxies and Tracers of Mass Accretion}

\author{F. Combes }
{ Observatoire de Paris, DEMIRM, Paris, France}

\begin{moriondabstract}
Can the present dynamics of spiral galaxies tell us something
about the merging history, the formation and evolution
of disks? Galaxy interactions thicken or destroy disks;
the simultaneous presence of thick and thin disks
is a tracer of past accretion, and also of recent
disk re-formation. Observation of a large 
number of counter-rotating disks is also
evidence of past mergers, as well as the frequency of
polar-ring galaxies. Finally, the ubiquitous presence
of warps in the outer parts of HI disks might also provide 
a clue of how frequently disks accrete mass with
different angular momentum.
\end{moriondabstract}

\section{Introduction}

According to hierarchical cosmological scenarios,
galaxies form through merging of smaller entities.
 At least the dark haloes of dissipationless matter
hierarchically merge, and it is expected that 
some of the visible galaxies also interact and
exchange mass, while spiraling in a common halo.
If major mergers lead to the formation of
ellipticals, and leave vestiges such 
as shells, ripples and loops around present-day
elliptical galaxies (Schweizer \& Seitzer 1988),
signatures of accretion or merger are less easy to
see in spiral galaxies. Yet, mass and
gas accretion is required in spiral 
galaxies for several reasons:

\begin{itemize}

\item  Metallicity distribution in the 
disk (the G-dwarf problem for instance, requires
gas infall)

\item Spiral formation and maintenance: episodes
of spiral density waves heat the disk,
and accreted fresh gas is required to trigger
new instabilities

\item Renewal of bars and  nuclear bars, that
drive mass towards the center, and self-destroy

\item Reforming the thin disk after minor mergers:
galaxies such as the Milky Way re-form a thin disk, 
while a thick one has been heated by an interacting event

\end{itemize}

In the following, the main evidences
for mass accretion in spiral galaxies will be reviewed, 
including: the presence of thick disks,
counter-rotating components, the ubiquity of
warps, or the existence of polar rings.

\section{Galaxy Interactions and
Thickness of Stellar Disks}

In hierarchical cosmologies, it is easy to estimate
analytically the probability of formation of
a dark halo of mass $M$ at time $t$
from the Press-Shechter theory (1974),
revised by Bond et al (1991): a gaussian distribution
of fluctuations is assumed, and structures are followed
through random walk of linear overdensity with
respect to smoothing scale.

From such analytical formulations of merging histories (e.g.
Lacey \& Cole 1993, 1994), it is possible to relate the dark haloes
 merger rate to the parameters of the universe (average density,
cosmological constant). The merging rates for visible galaxies should
follow, although the link is presently not well known
(Carlberg 1991, Toth \& Ostriker 1992).

For the standard CDM model ($\Omega$=1) for instance,
80\% of haloes have accreted 
at least 10\% of their mass in the last 5 10$^9$ yrs.
To reduce the merging rate today, the solution
is to consider low $\Omega$ models, 
for which freezing of halo formation occurs 
for z $<$ 1/$\Omega$. After this epoch,
only very few haloes form, and the merger
rate of visible galaxies inside haloes is expected
also very low. When approximations are taken
for the merger conditions of the galaxies,
such as a threshold in their relative velocity
v $< v_{mg} \sim v_{escape}$
(Carlberg 1990, 91), the merger rate can be written
as a power law with redshift, 
$dn(mergers) /dt  \propto  (1+z)^m $,
with the power-law $m$ increasing with $\Omega$
and $\Lambda$ (typically as 
$m \propto  \Omega^{0.42} (1-\Lambda)^{-0.11}$).

Observations
support a large value of the exponent $m$.
Statistics of close galaxy pairs from faint-galaxy redshift surveys
have shown that the merging rate increases as
 $(1+z)^{m}$ with $m=4\pm1.5$ (e.g. Yee \& Ellingson 1995). Lavery
et al (1996) claim that collisional ring galaxies
(Cartwheel-type) are also rapidly evolving,
with $m=4-5$, although statistics are still insufficient.
Many other surveys, including IRAS faint sources,
or quasars, have also revealed a high power-law (Carlberg 1991,
Carlberg et al 1994).

\bigskip

The fragility of disks with respect to interactions
can be used to constrain the merging rate.
During an interaction, stellar disks can 
thicken or even be destroyed (e.g. Gunn 1987).
Through the disk thickness of the Milky Way,
Toth \& Ostriker (1992) constrain the frequency of merging and the 
value of the cosmological parameters: from analytical and local
estimations of the heating rate, they claim that 
the Milky Way disk has accreted less than 4\% of its mass within the 
last 5 Gyrs. But these local calculations are only
rough approximations. The first 
numerical simulations of the phenomenon of disk
thickening through interactions (Quinn et al 1993, Walker et al 1996)
appear to confirm the analytical results however: they show 
that the stellar disk thickening can be large and sudden.

Recently, Huang \& Carlberg (1997) and Velazquez \& White (1999)
reconsider the problem, through numerical simulations,
 and find on the contrary that the
 heating of disks have been overestimated. In particular,
prograde satellites heat the disks, while 
retrograde ones produce only a coherent tilt.
If the halo is rigid, the thickening of the disk is 
increased by a factor 1.5 to 2: massive live bulges
can therefore help to keep disks thin, in absorbing
part of the heating.
Also, there are many parameters
to explore in simulations: the most important could be
the compactness of the interacting companion. If
the perturber has a compact core, the heating effect is important,
while a more diffuse  companion is destroyed by tidal
shear before damaging the primary disk. 

\bigskip

\begin{figure}
\psfig{figure=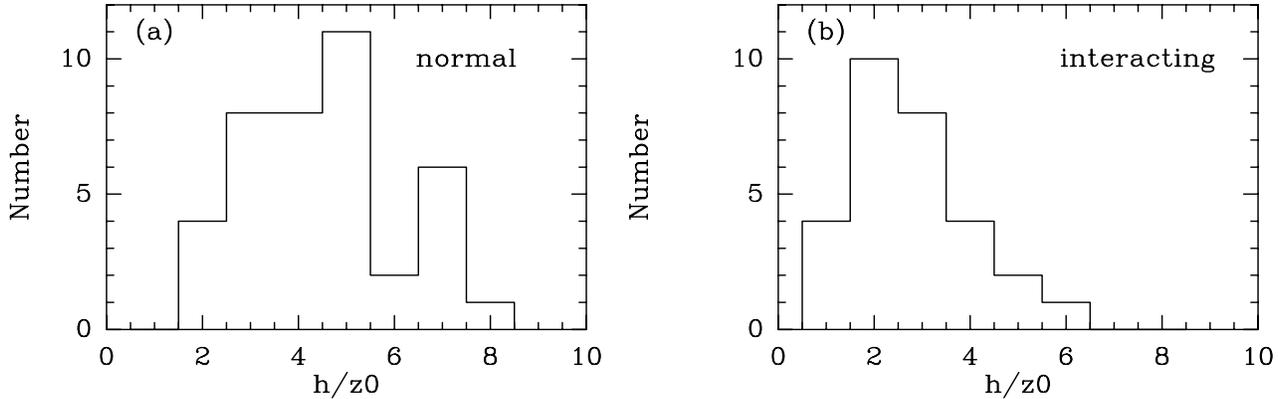,width=17cm,bbllx=3cm,bblly=0mm,bburx=12cm,bbury=25cm,angle=-90}
\caption{Distribution of scalelength to scale height ratios
for (a) normal galaxies and (b) interacting galaxies (from Reshetnikov \& Combes, 1997).}
\label{fig1}
\end{figure}

It should be however remarked that gas hydrodynamics and star 
formation processes can also alter significantly the
processes, since the thin disk can be reformed 
continuously through gas infall.
 It is interesting to check on presently interacting galaxies
whether the heating or thickening of disks is measurable.
 In normal galaxies, the ratio of radial scale-length $h$
to scale-height $z_0$ is about constant and equal to 5
(Bottema 1993); the ratio only goes up for dwarf galaxies.
Now, in a sample of edge-on interacting galaxies this ratio
was found to be 1.5 to to 2 times lower than normal (Reshetnikov \& 
Combes 1997, Schwarzkopf \& Dettmar 1999), as 
shown in fig \ref{fig1}. This is surprising,
when taking into account that the visible "interacting phase"
is only transient, and on a Gyr time-scale, interacting
galaxies will return to the "normal phase", with again 
a high $h/z_0$ ratio, or thin disk.

A possible interpretation of this result is that the interacting
galaxies are warped, since the latter is difficult to distinguish
from a thickening at any viewing angle; yet the damping of the warp 
will thicken the disk in any case. 
If the thickening is in fact transient, this indicates
that the present disk galaxies come from merging of smaller units,
that they acquire mass continuously: through gas accretion
and subsequent star formation,
disks recover their small thickness after galaxy interactions,
or in other words,
the disk of present day spirals has been  assembled at low 
redshift (Mo et al 1998).

\section {Counter-Rotating Components}

The phenomenon of gas disks counter-rotating with the
stellar component is now well known in ellipticals,
where the ionised gas disks (or dust lanes) are settled 
in principal planes (e.g. Bertola et al 1990).
Ellipticals were also first discovered with kinematically decoupled 
stellar cores, which are expected in merger remnants, such as NGC 7252
(e.g. Barnes \& Hernquist 1992).

Counter-rotation has also been observed in many spirals during
this last decade, although
it is more difficult than in ellipticals,
since the secondary CR component is not dominating
and the primary component is strongly rotating.
All possibilities have been observed, either two stellar disks
counter-rotating with respect to each-other, or the gas counter to the stars,
or even gas versus gas, but not at the same radii in the galaxies
(see the reviews of Galletta 1996, Bertola \& Corsini 1998).
 There is presently about 60 systems of counter-rotation
recorded in the literature. 

In general the counter-rotating component is not dominant, but
there is a very special case, NGC 4550, where 
two almost identical CR stellar disks are observed
(Rubin et al 1992). This case is a puzzle, since 
the second disk cannot have formed through subsequent 
accretion of gas: the two stellar disks have the same age. 
The only solution is through a merger of pre-existing
spiral galaxies. If major mergers usually give
an elliptical galaxy as a remnant, this is not the case
when they have aligned direction of their angular momentum.
In these rare orientation cases, it is possible to 
merge two spiral galaxies in one, and reproduce
the case of NGC 4550, when the momenta are opposite
(Pfenniger 1998, Puerari \& Pfenniger 1999).

Can one consider other explanations than mergers
for CR components? It is possible to artificially 
simulate counter-rotation in a certain
region of a galaxy, through perpendicular streaming
motions due to a bar potential, for instance;
but when the 2D velocity field is
obtained, confusion is not possible.  There are also
self-consistent models of barred galaxies including retrograde
orbits (Wozniak \& Pfenniger 1997), but the origin
of the retrograde stars is still gas accretion.
A slow bar destruction can be a rare case
where stars in box-orbits in a barred galaxy are
scattered equally in two CR families of tube orbits,
resulting in two opposite streams when the bar has 
disappeared (Evans \& Collett 1994). But the process
of bar destruction is in any case related to
galaxy interactions and gas accretion.

\subsection{Stability}

How long such counter-rotating systems can live?
 Does this phenomenon favor gas fueling to the 
nucleus?

There exists a two-stream instability in flat disks,
similar to that in CR plasmas (Lovelace et al 97).
If there exists a mode in a given disk, the
energy of the modes in the two
streams are of opposite signs: the negative E mode can 
grow by feeding energy in the positive E mode,
which produces the instability.
There exist also many bending instabilities
(Sellwood \& Merritt 1994). 

If there is only a small fraction of CR stars, these
haveon the contrary a stabilising influence with 
respect to bar formation ($m=2$); in a certain sense, they
are equivalent to a system with more velocity dispersion
(Kalnajs 1977).

\begin{figure}
\psfig{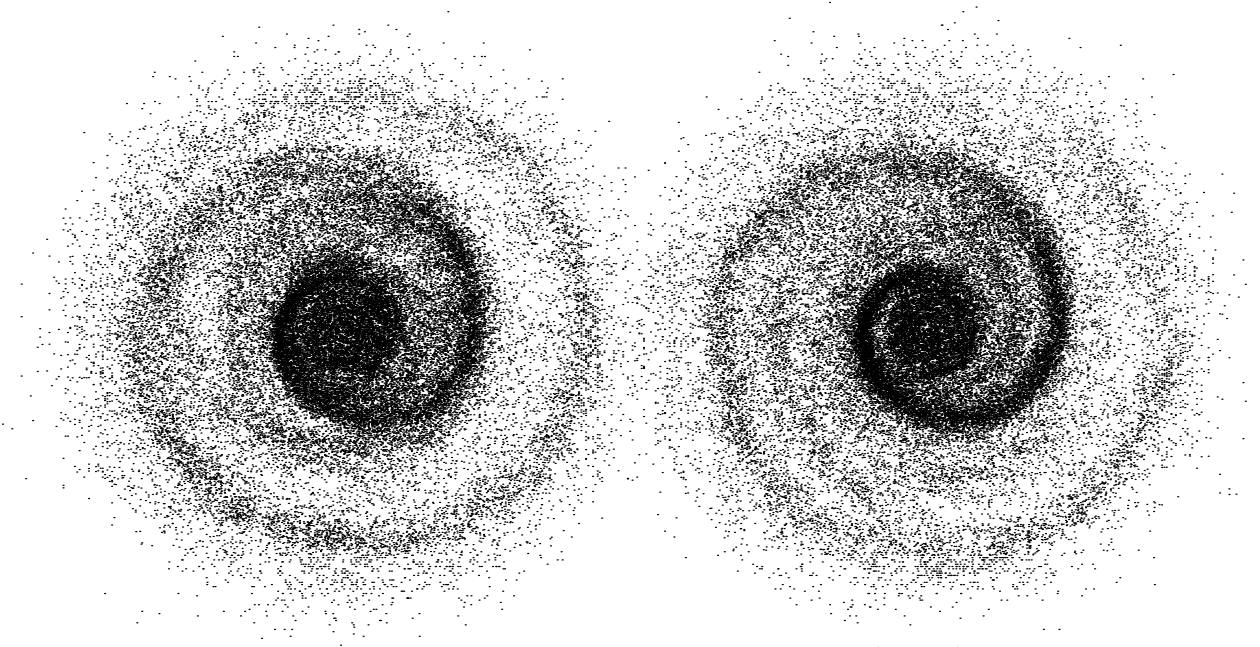}
\caption{N-body simulation of two coexisting disks of stars, showing
the $m=1$ instability, a leading arm with respect to the direct stars (left)
and trailing for the retrograde stars (right), here plotted separately.
The instability occurs only if the CR stars represent at least 25\%
of the total.}
\label{fig2}
\end{figure}

But in the case of comparable quantities of CR stars,
a one-arm instability is triggered. This is confirmed
through N-body simulations:
a quasi-stationary one-arm structure forms,
and lives for 1--5 periods
(Comins et al 1997), first leading, than trailing, and disappears.
Fig \ref{fig2} shows such a simulation, where the
common $m=1$ pattern is leading for the main direct component,
and trailing for the secondary retrograde one.

\subsection{ Counter-rotating gas}

Accretion of CR gas in a lenticular galaxy
deprived of gas initially is a way to form
two stellar counter-rotating disks, after
star formation has occured. 
Thakar \& Ryden (1996) have shown that
both episodic or continuous gas infall are
able to form a stable CR disk, without de-stabilising
the pre-existing disk significantly. The conditions
are that gas must be extended in phase space and not clumpy,
which would heat too much the primary disk. 
For example, the merger of a gas-rich dense dwarf will have 
a too large heating effect, unless the mass ratio is
quite small, and in such cases, only a small CR disk 
is produced. However, the final thickness of
the disk depends drastically on the gas code used,
thicker for sticky particles, and much thinner with SPH 
(Thakar \& Ryden 1998), as well as the settling time-scales.

When gas is present in the initial disk, 
the presence of two CR streams of gas in the same plane will 
be very transient: strong shocks will
produce heating and rapid dissipation will drive
the gas quickly to the center (Kuznetsov et al 1999).
This could be a very efficient way to fuel active
nuclei. However, the gas could also infall
in an inclined plane, or at different radii 
than those of the pre-existing gas, which can explain
the observations of two counter-rotating gas systems.

Polar rings (objects similar to the prototype NGC 4650A) 
are such cases, where gas settles in a stable plane almost 
perpendicular to the primary galaxy.
Polar-ring galaxies are quite rare in the nearby universe:
Whitmore et al (1990) find that about 0.5\% of all nearby
lenticular galaxies are observed with a polar ring. But since
there are projection effects and different selection biases
that prevent to see them all, they estimate to about 5\% the
actual present frequency of PRGs.
 An estimation of their frequency as a function of redshift
will be a precious tool to quantify the merging rate evolution.

\section{Warps as clues of matter accretion}

The majority of spirals are warped in their neutral hydrogen (HI)
component (e.g. Sancisi 1976, Bosma 1981, Briggs 1990).
This is a long-standing puzzle, since if the gas is considered 
as test-particles in the halo potential, it should
differentially precess, and with a time-scale much shorter
than the Hubble time the disk should end up with a 
corrugated shape and thicken.

Many theories have been proposed to solve the problem.
Normal modes of the disk have been ruled out, since they are
quickly damped (Hunter \& Toomre 1969), but
normal modes of the disk in the potential if
a mis-aligned halo have been
a possibility for a while (Toomre 1983, Sparke \& Casertano 1988),
until it was realized that they are
quickly damped through dynamical friction  (Nelson 
\& Tremaine 1995).

\begin{figure}
\psfig{figure=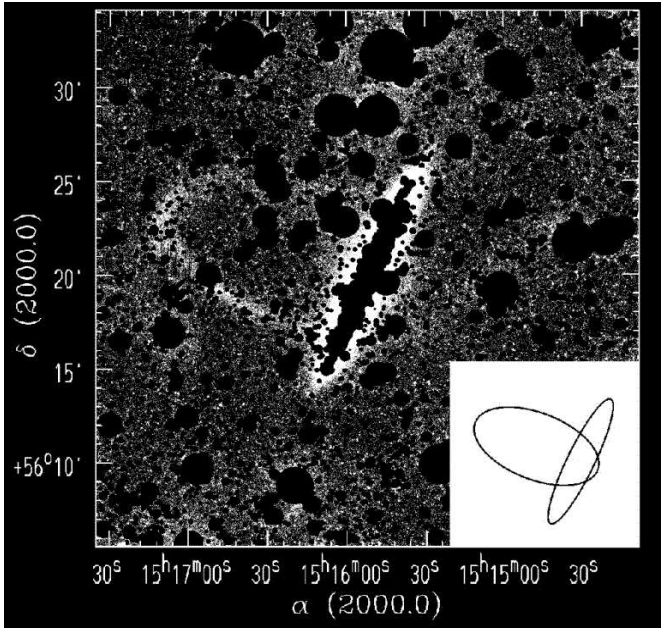,width=9cm,bbllx=1cm,bblly=5cm,bburx=20cm,bbury=24cm,angle=0}
\caption{NGC 5907 image obtained by Shang et al. (1998) in the visible
(6600\AA). All foreground stars and the dust lane of the edge-on
galaxy have been blanked. The inset diagram shows the position 
of the ring/loop (from Shang et al. 1998).}
\label{fig3}
\end{figure}

The triggering of warps by tidal interaction
with companions has been ruled out in the past, since 
the best examples of warped galaxies appeared isolated.
 However, this could be changing now that smallest companions
can be found, or vestiges of a past merger. This is the case
of the warp-prototype NGC 5907, where a conspicuous tidal loop
has been observed by Shang et al (1998). It is obvious that 
this galaxy has accreted a small system in the recent
past, and it has also a dwarf companion nearby
(see Fig \ref{fig3}).  
Regular and symmetric warps are those that have already
relaxed for a while, and this could explain the apparent lack
of correlation with companions.

Finally, the proposition that gas infall could
maintain warps around galaxies is easily justified
in the framework of hierachical cosmologies (cf Ostriker \& Binney
1989; Binney 1992). Gas infalls with slewed angular momentum
with respect to the main disk. This accretion will 
re-align the whole system along a tilted axis.
The transient state is the warped state. This hypothesis
has been recently supported through numerical
simulations by Jiang \& Binney (1999).
They show that the inner halo and disc tilts as one unit.
The halo tilts first in the outer parts, and the 
tilt propagates then inwards; the disk is entrained
and aligns with the halo, it plays the role of a tracer 
of its orientation. The time-scale of this phenomenon is
about 1 Gyr to re-align by 7$^\circ$.

\section {Conclusions}

In summary, there are many evidences that even spiral
galaxies have experienced a large number of
galaxy interactions in the past, and that their
formation proceeds also through hierachical merging
and accretion: presence of thick and thin disks,
growing number of observed counter-rotating disks,
frequency of polar-ring galaxies, ubiquity of HI warps.
Both present explanations of warps, either through 
tidal interactions trigger, or maintenance through gas infall,
are compatible with this paradigm.

\vfill

\begin{moriondbib}
\bibitem{} Barnes J., Hernquist L., 1992, {\it Ann. Rev. Astron. Astrophys.}
{\bf 30}, 705
\bibitem{}Bertola F. Bettoni D., Buson L.M., Zeilinger W.W., 1990, in 
{\it Dynamics and Interaction of Galaxies}, ed. R. Wielen, Springer, p. 249
\bibitem{}Bertola F. \& E. Corsini, 1998 in {\it Galaxy Interactions} 
IAU Symp. 186, Kyoto 1997, Kluwer
\bibitem{} Binney J., 1992, {\it Ann. Rev. Astron. Astrophys.}
{\bf 30}, 51
\bibitem{}Bond J.R., Kaiser N., Cole S., Efstathiou G., 1991, 
\apj {379}{440}
\bibitem{}Bottema R., 1993, \aa {275}{16}
\bibitem{}Bosma A., 1981, \aj {86}{1825}
\bibitem{}Briggs F., 1990, \apj {352}{15}
\bibitem{}Carlberg R.G. 1990, \apj {359}{L1}
\bibitem{}Carlberg R.G. 1991, \apj {375}{429}
\bibitem{}Carlberg R.G., Pritchet C.J., Infante L. 1994 \apj {435}{540}
\bibitem{}Comins N.F., Lovelace R.V.E., Zeltwanger T., Shorey P., 1997,
\ apj {484}{L33}
\bibitem{}Evans N.W., Collett J.L., 1994, \apj {420}{L67}
\bibitem{}Galletta G., 1996, in {\it Barred Galaxies}, ed R. Buta, D.A. Crocker
\& B.G. Elmegreen, ASP Conf. Series. {\bf 91}, p. 429
\bibitem{}Gunn J.E., 1987, in {\it Nearby Normal Galaxies} ed. S.M. Faber, 
Springer New York, p. 459
\bibitem{}Huang S. \& Carlberg R.G. 1997, \apj {480}{503}
\bibitem{}Hunter C., Toomre A., 1969 \apj {155}{547}
\bibitem{}Jiang I-G., Binney J., 1999, \mnras {303}{L7}
\bibitem{}Kalnajs A., 1977, \apj {212}{637}
\bibitem{}Kuznetsov O.A., Prokhorov M.E., Sazhin M.V., Chechetkin V.M.,
1999, {\it Astrophys. J.} preprint (astro-ph/9810429)
\bibitem{}Lacey C., Cole S. 1993, \mnras {262}{627}
\bibitem{}Lacey C., Cole S. 1994, \mnras {271}{676}
\bibitem{}Lavery R.J., Seitzer P., Suntzeff N.B., Walker A.R., 
Da Costa G.S. 1996, \apj {467}{L1}
\bibitem{}Lovelace R.V.E., Jore K.P., Haynes M.P., 1997, \apj {475}{83}
\bibitem{}Mo H.J., Mao S., White S.D.M., 1998, \mnras {295}{319}
\bibitem{}Nelson R.W., Tremaine S., 1995, \mnras {275}{897}
\bibitem{}Ostriker E.C., Binney J.J., 1989, \mnras {237}{785}
\bibitem{}Pfenniger D., 1998, in {\it Galaxy Interactions} IAU Symp. 186, Kyoto
1997, Kluwer
\bibitem{}Press W.H., Schechter P.  1974, \apj {193}{437}
\bibitem{}Puerari I., Pfenniger D., 1999, in {\it The Evolution of Galaxies 
on Cosmological Time-Scales}, Tenerife, Spain, Nov 30-Dec 5, 1998, PASP Series
ed T. Mahoney \& J.E. Beckman 
\bibitem{}Quinn P.J., Hernquist L., Fullagar D.P., 1993, \apj {403}{74}
\bibitem{}Reshetnikov V., Combes F.: 1997, \aa  {324}{80}
\bibitem{}Rubin V.C., Graham J.A., Kenney J.D.P., 1992, \apj {394}{L9}
\bibitem{}Sancisi R., 1976, \aa {53}{159}
\bibitem{}Schwarzkopf U., Dettmar R-J., 1999, in {\it Galaxy Evolution:
Connecting the Distant Universe with the Local Fossil Record}, ed. M.
Spite, Kluwer
\bibitem{}Schweizer F., Seitzer P.  1988, \apj {328}{88}
\bibitem{}Sellwood J.A., Merrit D., 1994, \apj {425}{530}
\bibitem{}Shang Z., Brinks E., Zheng Z. et al., 1998, \apj {504}{L23}
\bibitem{}Sparke L., Casertano S., 1988, \mnras {234}{873}
\bibitem{}Thakar A.R., Ryden B.S., 1996 \apj {461}{55}
\bibitem{}Thakar A.R., Ryden B.S., 1998, \apj {506}{93}
\bibitem{}Toomre A. 1983, in {\it Internal Kinematics and Dynamics of
Galaxies}, IAU Symp. 100, ed. E. Athanassoula, p. 177 (Reidel)
\bibitem{}Toth G., Ostriker J.P. 1992 \apj {389}{5}
\bibitem{}Velazquez H., White S.D.M., 1999, \mnras {304}{254}
\bibitem{}Whitmore B.C., Lucas R.A., McElroy D.B. et al: 1990, \aj {100}{1489}
\bibitem{}Yee H.K.C., Ellingson E. 1995 \apj {445}{37}
\bibitem{}Walker I.R., Mihos J.C., Hernquist L., 1996, \apj {460}{121}
\bibitem{}Wozniak H., Pfenniger D., 1997 \aa {317}{14}

\end{moriondbib}
\end{document}